\documentstyle[11pt,twoside,psfig,jltp]{article}

\title{Ferromagnetic Nanowires with Superconducting Electrodes}

\author{V. T. Petrashov\address{Department of Physics, Royal Holloway,
University of London, Egham, Surrey, TW20 0EX, U.K.}, I. A.
Sosnin, I. Cox, A.Parsons, C. Troadec}

\runninghead{V. T. Petrashov \it{et al.}}{Ferromagnetic Nanowires
with Superconducting Electrodes}

\begin{document}

\begin{abstract}
The proximity effect in mesoscopic ferromagnet/superconductor
($FS$) Ni/Al structures of various geometries was studied
experimentally on both $F$- and $S$-sides of the structures.
Samples with a wide range of interface transparency were
fabricated. The dependence of the effect on $FS$ interface
transparency was investigated. The amplitude of this effect was
found to be larger than expected from classical theory of
proximity effect. Preliminary experiments showed no
phase-sensitive oscillations in Andreev interferometer geometry.
Various theoretical models are discussed.

PACS numbers:74.50.+r, 74.80. Fp, 85.30. St.
\end{abstract}

\maketitle


\section{INTRODUCTION}
Electronic devices exploiting the spin of conduction electrons
rather than their charge have been proposed recently as an
alternative to conventional electronics (see e.g. Ref. 1 and
references therein). Ferromagnetic materials, being a natural
source of spin-polarized electrons for such devices, are in focus
of intensive experimental and theoretical investigations. Hybrid
ferromagnet/superconductor ($FS$) structures are prospective
candidates for device application and can be useful tools for
studying properties of nanometer-size ferromagnets. Recently the
measurements of the spin polarization of direct current have been
reported for ballistic point contacts \cite{Soulen,Upadhyay}.
These experiments were in reasonable agreement with both the band
structure of ferromagnets \cite{Stearns} and the general picture
of Andreev reflection on the $FS$ interface \cite{deJong}. In
contrast, recent experiments on diffusive $FS$ nanostructures
showed a dramatic disagreement between the theory and the data.
While the theory \cite{Demler} predicts any superconducting
correlation to decay in the diffusive ferromagnet over the
distance $\xi_{F}=\sqrt{\frac{\hbar D}{k_{B}T_{C}}}$ governed by
the exchange energy of the ferromagnet, which is of the order of
$k_{B}T_{C}$ ($T_{C}$ is the Curie temperature, $D$, the diffusion
constant of the ferromagnet), the experimental results suggest
that the influence of the superconductor penetrates into the
ferromagnet over a distance up to $10^{2}$ times larger than
$\xi_{F}$ \cite{Petrashov,Pannetier,Sosnin}.

Here we report further experimental studies of mesoscopic $FS$
structures of various geometries. We find that the conductance
changes can be of both negative and positive sign at the
superconducting transition, with the amplitude of the changes up
to $10^{2}$ times larger than theoretical values. We demonstrate
that the sign and the amplitude of the effect depend strongly on
the interface transparency. Our preliminary experiments with $FS$
interferometers showed no phase-periodic oscillations down to the
level of 0.1 $e^{2}/h$.

\section{EXPERIMENTAL}
The samples were fabricated using multiple e-beam lithography.
Ni/Al structures were thermally evaporated
\begin{figure}[ht]
\centerline{\psfig{file=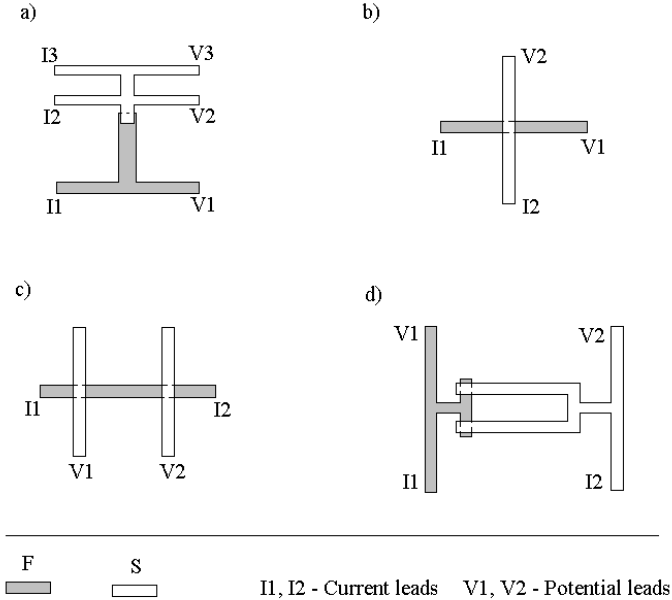,height=3.6in}} \caption{Sample
geometries} \label{Fig.1}
\end{figure}
in a vacuum of $10^{-6}$ mbar onto a silicon substrate kept at
room temperature. The first layer was a ferromagnet (Ni) 40nm
thick, the second layer, a superconductor (Al) 60 nm. Various
geometries studied are shown in Fig. 1. The resistivities of the
films varied from sample to sample and were in the range of 10 -
50 $\mu\Omega$ cm for Ni and 1.0 - 1.5 $\mu\Omega$ cm for Al.

The resistance of the structures was measured by the four-terminal
method. Current and potential leads are marked $I$ and $V$ in Fig.
1. The resistance of the structures was measured using both dc and
ac signals in the temperature range from 0.27K up to 50K and in
magnetic fields up to 5T.

Special care was taken to create interfaces of controllable
quality. Before the deposition of the second layer, the contact
area was Ar$^{+}$ plasma etched. By varying etching parameters, we
obtained interfaces of a wide range of transparencies. Wide
checking layers were analyzed by Secondary Ion Mass-Spectroscopy
(SIMS) with the primary beam of Cs$^{+}$ ions. For the best
samples, the concentrations of oxygen and carbon at the Ni/Al
interface were in the range of 0.1-0.01 of a single atomic layer.

\section{RESULTS}
Figures 2 a and b
\begin{figure}[p]
\centerline{\psfig{file=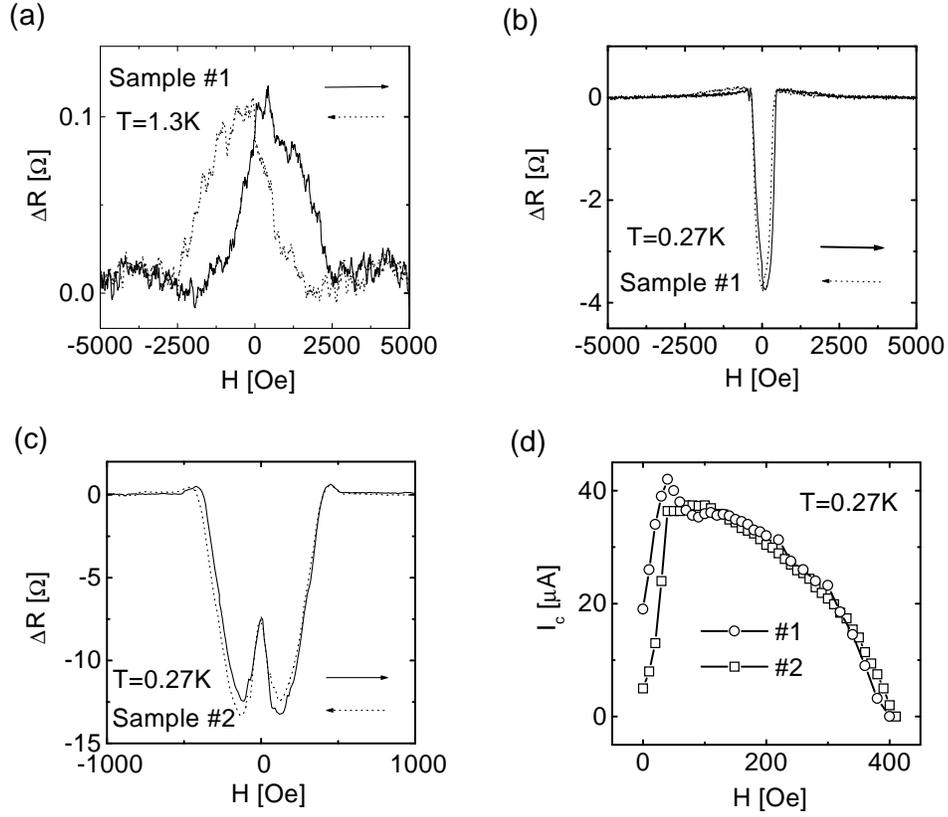,height=4.4in}}
\caption{Magnetoresistance of two samples in geometry Fig. 1a. a)
Sample $\#$1 ($I1$,$I2$,$V1$,$V2$) above the superconducting
transition at $T$=1.3K b) Sample $\#$1 ($I1$,$I2$,$V1$,$V2$) below
the transition at $T$=0.27K. c) Sample $\#$2 ($I1$,$I2$,$V1$,$V2$)
at $T$=0.27K. d) Critical current of the superconducting
transition of adjacent superconductor structure
($I2$,$I3$,$V2$,$V3$) for samples $\#$1 and $\#$2.} \label{Fig.2}
\end{figure}
\begin{figure}[p]
\centerline{\psfig{file=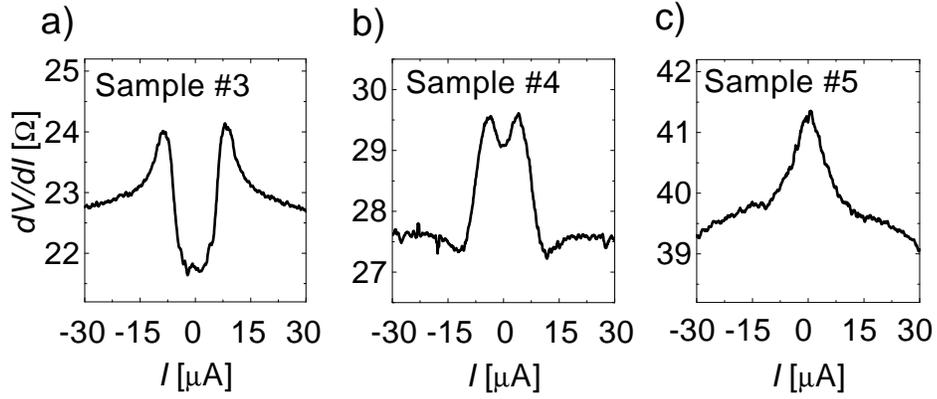,height=2.25in}}
\caption{Differential voltage-current characteristics of three
different samples in geometry of Fig. 1b. a) Sample $\#$3,
$R_{b}$=22.5$\Omega$; b) Sample $\#$4, $R_{b}$=28$\Omega$; c)
Sample $\#$5, $R_{b}$=39$\Omega$.} \label{Fig.3}
\end{figure}
\begin{figure}[p]
\centerline{\psfig{file=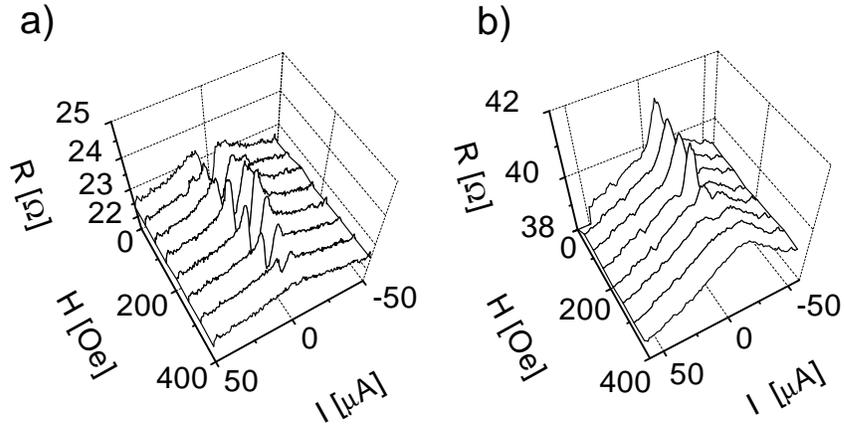,height=3in}} \caption{Applied
current - magnetic field 3D-diagrams for samples $\#$3 (a) and
$\#$5 (b).} \label{Fig.4}
\end{figure}
\begin{figure}[p]
\centerline{\psfig{file=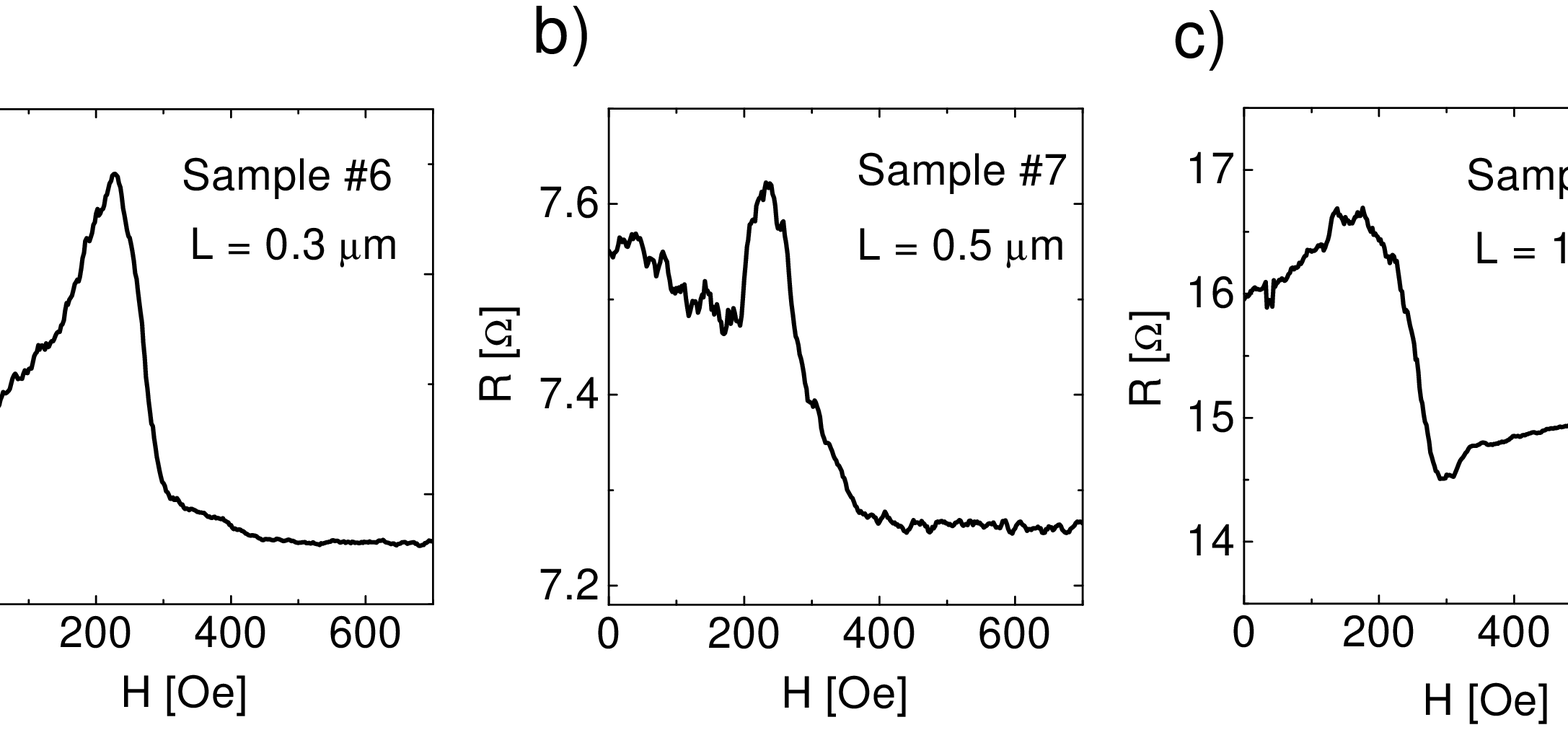,height=2.25in}}
\caption{Magnetoresistance of samples $\#$6-8 in geometry of Fig.
1c at $T$=0.27K. The length of the samples was a) 0.3$\mu$m, b)
0.5$\mu$m, and c) 1$\mu$m.} \label{Fig.5}
\end{figure}
\begin{figure}[p]
\centerline{\psfig{file=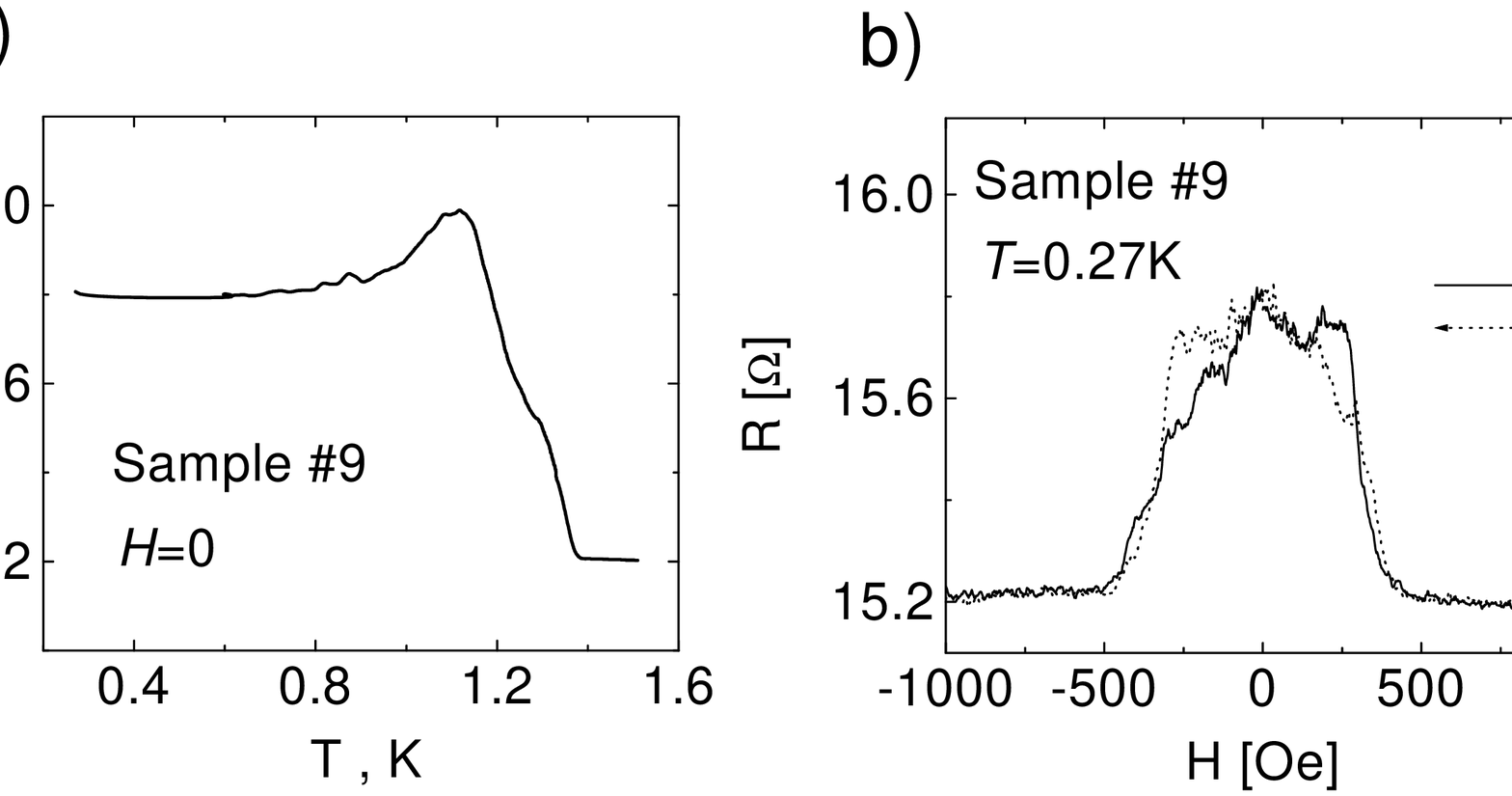,height=2.25in}}
\caption{Temperature dependence and magnetoresistance sample $\#$9
of the geometry of Fig. 1c with length $L$=1$\mu$m.} \label{Fig.6}
\end{figure}
show magnetoresistance of the $FS$ junction of sample $\#$1
(geometry of Fig. 1a) measured using contacts ($I1,I2,V1,V2$) at
temperatures well above ($T$=1.3K) and below ($T$=0.27K) the
superconducting transition. At $T$=1.3K the junction shows
anisotropy magnetoresistance with hysteresis typical for
ferromagnetic conductors. At temperatures below the transition,
the magnetic field dependence changes drastically. A resistance
drop is observed at the superconducting transition of Al in the
magnetic field. The amplitude of the drop is close to that
observed in the temperature dependence \cite{Sosnin}. The
hysteresial behaviour is also observed below the superconducting
transition.

Some samples showed negative magnetoresistance at small magnetic
fields (Fig. 2c). There is a correlation of this feature with the
critical current of the superconducting transition of the adjacent
superconducting structure ($I2,I3,V2,V3$) presented in Fig. 3d.
Note that the critical current for sample $\#$2 goes nearly to
zero at small magnetic fields, while that for the sample $\#$1
stays relatively large (about 20$\mu$A). The critical temperature
of the superconducting transition of the structure measured using
$I2,I3,V2,V3$ at zero magnetic field was within the range 1.0 -
1.05 K for all samples of this geometry.

To study properties of the $FS$ interface itself we used a cross
geometry shown in Fig. 1b. In Fig. 3 we show differential
voltage-current characteristics of three interfaces with different
transparencies. We see a cross-over from negative to positive
change in the resistance of the interface versus applied current
upon increase of the interface resistance. The cross-over takes
place at the specific interface resistance about 3x$10^{-9}$
$\Omega$cm$^{2}$. Temperature dependencies reflect the same
tendency \cite{Sosnin}.

Figures 4 a) and b) show 3D applied current - magnetic field
diagrams of samples $\#$ 3 and $\#$5 respectively. There are a
number of magnetic field independent peaks seen on these 3D
diagrams. These peaks are small in amplitude but are clearly seen
on all measured 3D diagrams of the contacts, including our best
ones with interface resistance below 1$\Omega$.

Proximity effect measured on samples with intermediate values of
the interface transparency (20$\Omega < R_{b} < 50\Omega$) showed
increase in the resistance upon the superconducting transition in
the geometry of Fig. 1c. Figure 5 shows magnetoresistance of three
samples of different length, $L$. All three show non-monotonic
magnetoresistance with maxima in resistance at a magnetic field
about 200 Oe. The final changes in the resistance,
$R(H=0)-R(H=700Oe)$, are larger for longer samples.

Figure 6 shows temperature dependence and magnetoresistance curves
for sample $\#$9, geometry Fig. 1c. The temperature dependence
shows a peak in resistance at the onset of superconductivity. The
magnetoresistance curve shows a hysteresis in superconducting
state similar to that of Fig. 2, but with opposite sign of
resistance changes. The magnetoresistance of this structure above
the transition showed the usual negative magnetoresistance of
small ($\Delta R = 0.05 \Omega$) amplitude.

We have also studied the electron transport in the Andreev
interferometer geometry of Fig. 1d. The magnetoresistance of the
structure was measured with an accuracy down to $\Delta
R/R^{2}\sim$ 0.1 $e^{2}/h$. In the first two samples tested, no
phase-sensitive oscillations were detected so far.

\section{DISCUSSION}
Our experimental data confirms the existence of long-range effects
in mesoscopic ferromagnet/superconductor structures, which was
established in earlier works
\cite{Petrashov,Giordano,Pannetier,Sosnin}. The changes in
conductance exceed greatly the value of $e^{2}/h$ which excludes
the mesoscopic origin of the effect observed. Such a giant
amplitude of the changes in conductance is yet to be explained.

There were several attempts to account for the long-range
superconducting proximity effect in ferromagnets. The authors of
Ref. 11 suggest that due to spin-orbit interaction in the
superconductor, the superconducting wave function may have a
triplet component. The lifetime of the triplet state in the
ferromagnet is much larger than that of a singlet one, therefore
this mechanism leads to the long-range effect. However, the
estimate based on the formulas presented in Ref. 11 gives the
values of the relative changes in conductance, $\Delta G / G$,
more than $10^{2}$ times smaller than the experimental values of
few percent.

The contribution of the interface to the conductance of hybrid
$FS$ structures has been addressed theoretically in both ballistic
\cite{deJong,Zutic} and diffusive \cite{Falko,Jedema,Golubov}
cases. The resistance of the diffusive $FS$ interface was
predicted to be always larger than that of the corresponding $FN$
one. In contrast, in Fig. 3 we see a decrease in the resistance of
the $FS$ interface for samples with higher interface
transparencies. For any interface transparency, we estimate the
effect of the shunting by the small part of the superconductor to
be of the order of or less than the resistance of one square of Al
film. In our case it is 0.1 - 0.3 $\Omega$. Since the changes in
the resistance in Fig. 3 are considerably larger, we believe that
shunting cannot explain the difference in behaviour.

The cross-over from positive changes in resistance to negative
ones presented in Fig. 3 can be accounted for using the
phenomenological analysis of Ref. 10. According to the latter the
changes in the resistance of the ferromagnetic wire, $\Delta
R_{FS}=R_{FN}-R_{FS}$, upon the superconducting transition can be
written as \cite{Sosnin}:

\begin{equation}
\frac{\Delta R_{FS}}{R_{FN}}=1-\frac{1}{\eta(1-P(1-\alpha))},
\end{equation}
where $P$ is the spin polarization and $\eta$ and $\alpha$ are
phenomenological parameters. $\eta$ is responsible for the
conductance enhancement due to Andreev reflection and varies in
the range $1\leq\eta\leq2$. Parameter $\alpha$ is proportional to
the amount of the spin polarized current in proximity to the
superconductor and varies in the range $0\leq\alpha\leq1$. Case
$\alpha = 0$ corresponds to total spin filtering (no spin
polarized current in the proximity of the ferromagnetic wire).
While the Andreev reflections increase the conductance of the $FS$
structure, the spin filtering decreases it. The competition
between the two determines the final sign of the conductance
changes. The two contributions may have different energy and
magnetic field dependencies which may lead to non-monotonic
dependencies like the ones presented in Fig. 5.

Magnetic field independent peaks on the $dV/dI$ versus current,
magnetic field 3D diagrams (Fig. 4), are present on all
measurements of the structures in the geometry of Fig. 1b. The
nature of the peaks is unclear.

The peak in resistance near the superconducting transition seen in
Fig. 6a may be explained by the charge imbalance effect caused by
the penetration of electric field into the superconductor
\cite{Golubov}. This resistance anomaly near the superconducting
transition has been suggested to be a measure of the spin
polarization of the ferromagnet \cite{Golubov} and it requires
additional experimental study.

The interesting feature of the presented data is the hysteresis in
the magnetoresistance of the $FS$ junctions. It is seen in Fig. 2
b) and c) and in Fig. 6 b). Note that the sign of the effect is
opposite but hysteresis is present in both cases. This effect
needs further investigation.

\section{CONCLUSIONS}
In conclusion, we have presented a systematic experimental study
of mesoscopic ferromagnet/superconductor structures of various
geometries. At the structures with high interface transparency we
measured a giant long-range proximity effect. Values of the
interface transparencies at which the cross-over from positive to
negative changes in the resistance at the superconducting
transition were found. Different theoretical approaches were
discussed. While the theory can qualitatively explain the
long-range superconducting proximity effect it fails to account
for the amplitude of the effect. Further experimental and
theoretical work is required.

\section*{ACKNOWLEDGMENTS}
We thank C. Lambert, A. Volkov, A. Golubov, A. Zagoskin, and V.
Chandrasekhar for useful discussions and A.F. Vyatkin for the SIMS
analysis of the samples. We appreciate technical support from M.
Venti. Financial support from EPSRC GR/L94611 is acknowledged.

\end{document}